\documentclass[aps,prl,twocolumn,showpacs,groupedaddress]{revtex4}  
\usepackage{graphicx}  
\usepackage{dcolumn}   
\usepackage{bm}        
\usepackage{amssymb}   
\usepackage{url} 

\begin{document}

\def\ncand{two}                          
\def\xbkg{1.24}                          
\def\ebkg{0.08}                          
\def\BR95{$1.2\times 10^{-7}$}           
\def\Dcut{0.949}                         

\def\D0{D0}                             
\def\met{\mbox{${\hbox{$E$\kern-0.6em\lower-.1ex\hbox{/}}}_T$ }} 
\def\gevc{${\rm GeV}/c$}                 
\def\gevcc{${\rm GeV}/c^2$}              
\def\ipb{${\rm pb}^{-1}$}                
\def\ifb{${\rm fb}^{-1}$}                
\def\bs2mm{$B^0_s \to \mu^+ \mu^-$}        
\def\bd2mm{$B^0_d \to \mu^+ \mu^-$}        
\def\mmm{$m_{\mu \mu}$}                  
\def\jpsi{J\!/\!\psi}                    
\def\bu2psik{$B^+ \to \jpsi K^+$}        
\def\Bbs2mm{${\mathcal B}(B^0_s \to \mu^+ \mu^-)$}    
\def\Bbd2mm{${\mathcal B}(B^0_d \to \mu^+ \mu^-)$}    
\def\psik{$\jpsi K^+$}                   
\def\psiphi{$\jpsi \phi$}                


\hspace{5.2in}\mbox{FERMILAB-PUB-07-395-E}

\title{Search for $B^0_s \to \mu^+\mu^-$ decays at \D0} 
%
\author{V.M.~Abazov$^{35}$}
\author{B.~Abbott$^{75}$}
\author{M.~Abolins$^{65}$}
\author{B.S.~Acharya$^{28}$}
\author{M.~Adams$^{51}$}
\author{T.~Adams$^{49}$}
\author{E.~Aguilo$^{5}$}
\author{S.H.~Ahn$^{30}$}
\author{M.~Ahsan$^{59}$}
\author{G.D.~Alexeev$^{35}$}
% ...

\affiliation{\vspace{0.1 in}(The D\O\ Collaboration)\vspace{0.1 in}}
% ...
\affiliation{$^{5}$University of Alberta, Edmonton, Alberta, Canada,
                Simon Fraser University, Burnaby, British Columbia, Canada,
                York University, Toronto, Ontario, Canada, and
                McGill University, Montreal, Quebec, Canada}
% ...
\affiliation{$^{28}$Tata Institute of Fundamental Research, Mumbai, India}
% ...
\affiliation{$^{30}$Korea Detector Laboratory, Korea University, Seoul, Korea}
% ...
\affiliation{$^{35}$Joint Institute for Nuclear Research, Dubna, Russia}
% ...
\affiliation{$^{49}$Florida State University, Tallahassee, Florida 32306, USA}
% ...
\affiliation{$^{51}$University of Illinois at Chicago,
                Chicago, Illinois 60607, USA}
% ...
\affiliation{$^{59}$Kansas State University, Manhattan, Kansas 66506, USA}
% ...
\affiliation{$^{65}$Michigan State University,
                East Lansing, Michigan 48824, USA}
% ...
\affiliation{$^{75}$University of Oklahoma, Norman, Oklahoma 73019, USA}
% ...
\date{July 26, 2007}

\begin{abstract}
We report
results from a search for the decay \bs2mm\ using 1.3
\ifb\ of $p\overline p$ collisions at $\sqrt{s}= 1.96$ TeV collected
by the \D0 experiment at the Fermilab Tevatron Collider.
We find
two candidate events, consistent with the expected background 
of $1.24 \pm 0.99$,
and set
an upper limit on the branching fraction of \Bbs2mm\ $<$ \BR95\ at 
the 95\% C.L.
\end{abstract}

\maketitle 

The branching fraction
\Bbs2mm\ 
is predicted to be
$(3.4 \pm 0.5) \times 10^{-9}$ \cite{Buras:2003td}
within the standard model (SM),
where
the decay occurs through helicity and CKM-suppressed processes 
involving multiple electroweak boson exchanges.
In supersymmetric (SUSY) models,
interactions with neutral Higgs bosons can enhance the branching ratio
by several orders of magnitude 
if the value of $\tan\beta$,
the ratio of vacuum expectation values for the two neutral CP-even Higgs fields,
is high 
\cite{
Babu:1999hn,
Choudhury:1998ze,
Dermisek:2003vn,
Blazek:2003hv,
Auto:2003ys}.
Large enhancements to ${\mathcal B}(B^0_s \to \mu^+ \mu^-)$ are possible in SUSY models
with $R$-parity violating couplings even if $\tan\beta$ is 
low \cite{Arnowitt:2002cq}.
Improvements to the limit on \Bbs2mm\ will constrain the parameter
space of such models.
The best published experimental bound is
${\mathcal B}(B^0_s \to \mu^+ \mu^-) < 2.0 \times 10^{-7}$ 
at the 95\% C.L. \cite{cdflim}.
The analysis reported in this letter used 1.3 \ifb\ of $p\overline p$ 
collisions collected by the \D0\ experiment at the Fermilab Tevatron. 
It supercedes our previous result \cite{Abazov:2004dj} based on 
a 240 \ipb\ subsample of the data.

The \D0\ detector~\cite{run2det} features
a three layer muon system~\cite{run2muon}
with each layer consisting of a scintillator plane
and a three or four plane drift chamber,
providing coverage for $\eta < |2|$,
where $\eta = -\ln[\tan(\theta/2)]$, and $\theta$
is the polar angle with respect to the beamline. 
Muon backgrounds are low due to shielding from 1.8 T iron
toroids located between the first and second muon detector layers,
and from a 6--10 interaction length deep uranium/liquid-argon calorimeter
located in front of the first layer.
Charged particles are detected in the inner central tracking system, which
consists of a silicon microstrip tracker (SMT) and a central fiber tracker
(CFT), both located within a 2 T superconducting solenoidal magnet. 
The CFT has eight thin coaxial barrels, each
supporting two doublets of overlapping scintillating fibers of 0.835~mm
diameter, one doublet being parallel to the beam axis, and the
other alternating by $\pm 3^{\circ}$. 
The SMT has four layers of double sided detectors divided into six
longitudinal sections interspersed with sixteen radial disks.
Each layer has a side with strips parallel to the beam axis; two layers
have a $\pm 2^{\circ}$ stereo side, and two layers have a $90^{\circ}$ side.
Typical strip pitch is $50-80$ $\mu$m.

Events were recorded using a set of single muon triggers, 
dimuon triggers, 
and 
triggers that selected $p\overline p$ interactions based on energy 
depositions in the calorimeter.
$B^0_s \to \mu^+\mu^-$ \cite{conj} candidates were formed from pairs of
oppositely charged muons.
Each muon was required to have transverse momentum $p_T > 2.5$ GeV, and 
to have hits in at least two layers of the muon system,
four layers of the CFT, and three layers of the SMT.
The $B^0_s$ candidate was required to have $p_T > 5$ GeV.
There is a large background due primarily to muons from the decay
of pions, kaons, and $b$- or $c$- flavored hadrons.
The $B^0_s \to \mu^+\mu^-$ signal is characterized by the long
lifetime of the $B^0_s$, which results in
an observable distance between the point at which the $B^0_s$ is
produced (the primary vertex) and the point at which it decays.
The distance from the primary vertex to the $B^0_s$ vertex
in the transverse plane ($L_T$) was required to
have an uncertainty $\sigma_{L_T} < 0.015$ cm and
a significance $L_T/\sigma_{L_T} > 12$.
The average $L_T$ for signal events passing the $p_T$ requirement
is $\sim 0.1$ cm.
Typically $\sigma_{L_T}$ is between 0.002 and 0.009 cm for both
signal and background.
The angle between the projections onto the transverse plane of 
the $B^0_s$ momentum 
and the displacement 
from the primary vertex to the $B^0_s$ vertex 
was required to be less than $15^{\circ}$.
The distance of closest approach $\delta$ of each muon to the primary vertex
in the transverse plane was calculated, along with the corresponding
uncertainty $\sigma_\delta$ and significance $\delta/\sigma_\delta$.
The smaller of the two significances, $\min(\delta/\sigma_\delta)$,
was required to be greater than 2.8.
This removes a class of events in which one of the tracks is consistent
with originating from the primary vertex.
A constrained fit was applied, enforcing the conditions 
that the tracks making up the
$B^0_s$ intersect in space and the three dimensional $B^0_s$ 
trajectory pass through the primary vertex.
The fit probability $P(\chi^2)$ is the fraction of the area of the $\chi^2$ 
distribution that lies below the $\chi^2$ value returned by the 
constrained fit.
It was required to be at least 0.01.

To further suppress the background, a likelihood ratio test was applied.
Five variables were incorporated: 
\begin{enumerate}
\item isolation, defined as $p_T^B/(p_T^B + \sum p_T)$
where $p_T^B$ is the transverse momentum of the $B^0_s$ system, and
$\sum p_T$ is the scalar sum of the transverse momenta
of all other tracks within a cone of $\Delta R < 1$ around the 
$B^0_s$ system, where $\Delta R = \sqrt{(\Delta \phi)^2 + (\Delta \eta)^2)}$
and $\phi$ is the azimuthal angle
\item $P(\chi^2)$
\item $L_{T}/\sigma_{L_T}$
\item $\min(\delta/\sigma_\delta)$
\item \mmm, the mass of the dimuon system.
\end{enumerate}
The likelihood ratio was approximated as $r = \prod_{i=1}^{5} S_i/B_i$
where $S_i$ is the probability distribution of the $i$th variable for the
signal, 
and $B_i$ is the distribution for the background.
The discriminant $D_5=r/(1+r)$ takes a value between zero (background-like)
and one (signal-like).
Figure \ref{fig-like_pdfs} shows the distributions of $S_i$
and $B_i$ for isolation and for functions of 
$L_{T}/\sigma_{L_T}$, 
$\min(\delta/\sigma_\delta)$,
and $P(\chi^2)$.
The functions 
map the quantities into the range zero to one.
They are given by 
$f_1(L_{T}/\sigma_{L_T}) = 1 - \exp[-0.057 (L_{T}/\sigma_{L_T} - 12)]$,
$f_2[\min(\delta/\sigma_\delta)] = 1 - \exp[-0.093(\min(\delta/\sigma_\delta) - 2.8)]$, and
$f_3[P(\chi^2)] = (P(\chi^2) - 0.01)/0.99$.
In Fig.\ \ref{fig-like_pdfs}, the signal and background events satisfy
all of the preselection cuts defined earlier except for the cut on
$L_T$ significance. To increase the statistics, the $L_T$ 
significance cut was relaxed from twelve to five. 
The signal distributions $S_i$ are given by the histograms in 
Fig.\ \ref{fig-like_pdfs}. 
These distributions are the result of Monte Carlo (MC) 
simulations using the {\sc pythia} event generator~\cite{Sjostrand:2000wi}
interfaced with the {\sc evtgen} decay package~\cite{evtgen},
followed by full {\sc geant} v3.15~\cite{geant}
modeling of the detector response. 
The simulation was tuned to reproduce the momentum resolution and scale, 
the trigger efficiency, 
and the $B^+$ meson $p_T$ distribution observed in data.
The MC events were processed with the same event reconstruction used for 
the data.
The background distributions $B_i$ are given by parameterizations of
the sideband data, shown in Fig.\ \ref{fig-like_pdfs}. 
The sideband data consist of candidates
having a dimuon invariant mass \mmm\ between 4.5 and 7.0 GeV excluding the 
signal
region. The signal region is between 4.972 and 5.717 GeV, approximately 
$\pm 3$ standard deviations around the mean of the Gaussian \mmm\
distribution in the signal MC. 
The sideband isolation distribution was fit to a Gaussian function, and the
other three sideband distributions were fit to the sum of two exponential
functions.
The \mmm\ distribution
of the background was approximated to be flat when computing the likelihood
ratio. 
The distribution of $D_5$ for signal and background is shown 
in Fig.\ \ref{fig-D5}.
Final candidates were required to have \mmm\ within the signal
region and to satisfy $D_5 > \Dcut$.
This threshold was chosen to optimize 
the expected 95\% C.L. upper bound on \Bbs2mm.
Two candidates pass the final selection.

\begin{figure}
\includegraphics[width=0.95\linewidth]{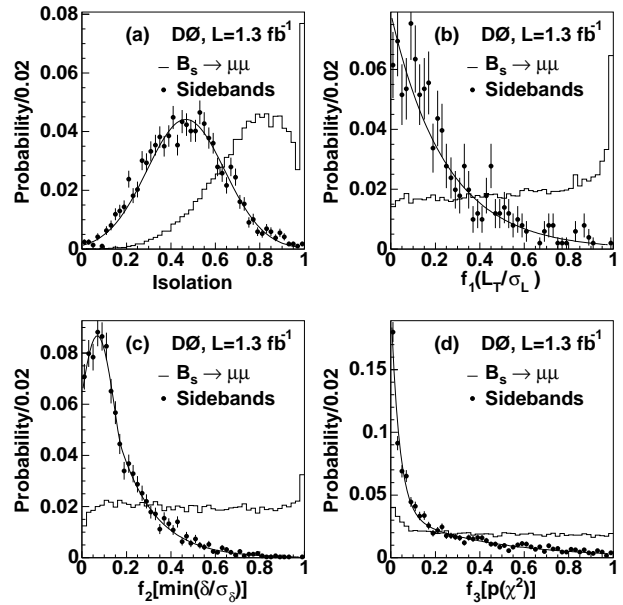}
\caption{\label{fig-like_pdfs} Signal and background distributions
  for four of the variables used in the likelihood ratio test.
  The signal distributions are from MC, and the background distributions
  are from the sideband data.
  The sideband distribution in (a) is parameterized as a Gaussian function.
  In (b), (c), and (d),
  the sideband distributions are parameterized as the sum of two
  exponential functions.
} 
\end{figure}
\begin{figure}
\includegraphics[width=0.95\linewidth]{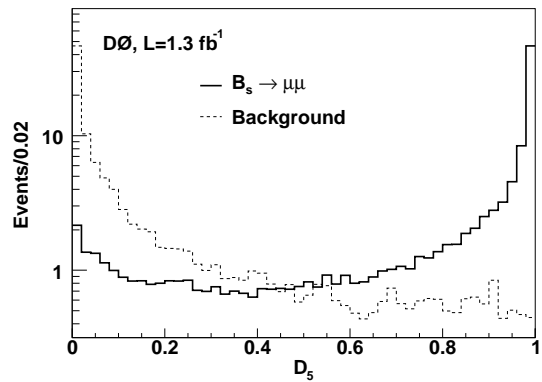}
\caption{\label{fig-D5} The distribution of discriminant $D_5$ for
  signal and background. 
  The background distribution is derived from events in the sidebands,
  folded over possible values of \mmm\ in the signal region.
  The signal distribution is from MC. The normalization 
  of the MC is arbitrary.
}
\end{figure}

An important feature of the background is seen in Fig.~\ref{fig-mmm},
which shows 
the distribution of \mmm\ after various cuts, 
beginning with the $L_T$ significance cut and ending
with $D_4 > 0.949$.
\begin{figure}
\includegraphics[width=0.95\linewidth]{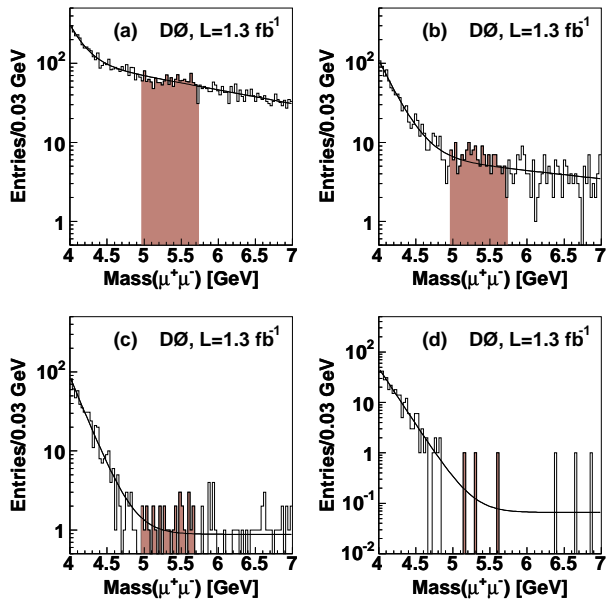}
\caption{\label{fig-mmm} The dimuon mass distribution at different
  stages in a sequence of cuts: 
  (a) after $L_T > 12\sigma_{L_T}$,
  (b) after $\min(\delta/\sigma_\delta) > 2.8$ and $P(\chi^2) > 0.01$,
  (c) after $D_4 > 0.5$, and (d) after $D_4 > 0.949$.
  In (a) and (b)
  the histograms are fit to the sum of two exponential functions,
  while in (c) and (d)
  they are fit to the sum of an exponential and a constant.
  The signal region (shaded) was excluded in the fits.
  In (d), three entries are included
  in the signal region: the entry near the upper bound of the signal region
  has a value of $D_4$ close to the threshold and fails the $D_5$ cut;
  the other two entries are the final candidates.
}
\end{figure}
The discriminant $D_4$ was calculated in the same way 
as $D_5$ except that the variable \mmm\ was omitted, thereby simulating 
the effect of a cut on $D_5$ without biasing the \mmm\ distribution
toward the $B^0_s$ mass.
Two components are evident in the distributions: 
a steeply falling component in the low mass region and a gradually falling
component whose slope diminishes as the cuts tighten.
This structure was studied using  $b\overline b$ 
events generated with {\sc pythia}, which reproduced the
main features of the data.
The contributions from particles misidentified as muons 
and other sources of real muons are small.
The gradually falling component consists of 
events in which the two muons arise from the decay of separate $b$ quarks,
while the steeply falling component consists of events 
in which the two muons arise from decay of the same $b$ quark,
via sequential decay $b\to c\mu\nu$ followed by $c\to s\mu\nu$ or
from $b\to\psi'X$ with $\psi'\to\mu\mu$. 
Higher mass $\psi'$ states may also contribute in the data.
Because the same-$b$ processes result from a single
$b$ quark, they have a better chance of producing a dimuon system 
that forms a common vertex and points back to the primary vertex 
than do the separate-$b$ processes.

The expected number of background events in the final candidate sample 
was estimated using events from the data in the low and high sidebands, 
together with the assumption that
the background consists of same-$b$ events having an
exponential mass distribution and separate-$b$ events having a 
flat mass distribution. 
This model of the shape of the backgrounds fits the 
sideband regions well and accurately predicts the number of events
in the signal region, see Fig.\ \ref{fig-mmm}.
The slope of the exponential 
was taken from the fit in Fig.\ \ref{fig-mmm}(d). 
The fits in Figs.\ \ref{fig-mmm}(c) and (d) are consistent
with a flat distribution for separate-$b$ events. 
The separate-$b$ distribution might still decrease gradually with mass 
after a cut on $D_4$, 
but the slope is not well constrained by the statistics in the high sideband, 
and to neglect it is conservative in its effect 
on the branching fraction limit.
Given the number of events in the low sideband and the slope of
the exponential, the expected contribution of same-$b$ events to 
the high sideband is negligible.
The estimated background from separate-$b$ events is 
$\sum_i P_i \cdot w$
where the sum is over all events in the high sideband.
The variable 
$w$ is the expected number of separate-$b$ events in the 
signal region per separate-$b$ event in the high sideband,
determined from the range of the signal region, the range of
the high sideband region, and the shape of the mass distribution
for separate-$b$ events.
The variable
$P_i$ is the probability for a separate-$b$ event to pass the cut
$D_5 > 0.949$ given that it falls within the signal region and
has the specific value of $D_4$ observed for the $i$th event
in the high sideband.
This probability was determined by 
integrating over the possible
mass values in the signal region. 
Likewise, the background from same-$b$ events was
computed using the corresponding sum over events in the 
low sideband.  
However, the low sideband contains separate-$b$ events as well as same-$b$
events.  As a result, the low sideband sum is an overestimate of the 
same-$b$ background.
The contribution due to separate-$b$ events in the low sideband was
estimated using the high sideband data and subtracted.
The total estimated background is $\xbkg \pm 0.99 \pm \ebkg$ events, where
the first uncertainty is statistical and 
the second is due to the uncertainty in the shape of the \mmm\ distribution.

The branching fraction was obtained by normalizing to the number of 
$B^+ \to \jpsi K^+ \to \mu^+\mu^- K^+$ candidates observed in the data. 
$B^+$ candidates were formed in a similar fashion to the $B^0_s$ candidates,
but with the addition of a third track, which was assumed to be
a kaon and required to have $p_T > 1.0$ GeV.
The three tracks had to form a common vertex,
and the two muons had to have a mass near the $\jpsi$ mass. 
As with $B^0_s$ candidates, 
the muon pair was required to have $\min(\delta/\sigma_\delta) > 2.8$. 
The $B^+$ system had to pass the same
$p_T$, angle, $\sigma_{L_T}$, $L_T/\sigma_{L_T}$, and $P(\chi^2)$ cuts
as the $B^0_s$ system.
Finally, the $B^+$ system was required to have $D_4 > \Dcut$.
The number of $B^+$ decays
$n_{B^+} = 2016 \pm 55 \, {\rm (stat)} \pm 45 \, {\rm (syst)}$ 
was determined from
the fit to the reconstructed mass distribution shown in 
Fig.\ \ref{fig-bufit}. 
\begin{figure}
\includegraphics[width=0.95\linewidth]{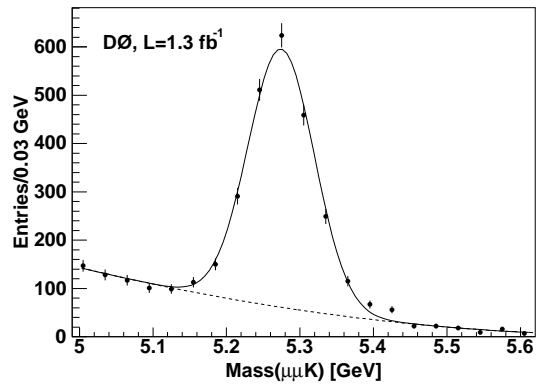}
\caption{\label{fig-bufit} {Mass distribution of $B^+$ candidates.
  The background distribution is parameterized as a parabola
  and the signal distribution as a Gaussian function.
  }
}
\end{figure}

The branching fraction is related to $n_{B^+}$ by 
\begin{eqnarray}
\label{eq-k}
{\mathcal B}(B^0_s\to\mu^+\mu^-) & = & \frac{n_{B^0_s}}{n_{B^+}}
             \cdot   \frac{\epsilon_{B^+}}{\epsilon_{B^0_s}}
             \cdot   \frac{f(\overline{b}\to B^+)}{f(\overline{b}\to B^0_s)}
                      \\ \nonumber
          & \times &  {\mathcal B}(B^+ \to \jpsi K^+) \cdot {\mathcal B}(\jpsi \to \mu^+\mu^-),
\end{eqnarray}
which is obtained by eliminating the integrated luminosity and
$b$ quark production cross section from the expressions for
the $B^+$ and $B^0_s$ yields.
The quantity
$n_{B^0_s}$ is the number of $B^0_s\to\mu^+\mu^-$ decays observed in the data.
The efficiencies $\epsilon_{B^+}$ and $\epsilon_{B^0_s}$ are, respectively, the
fractions 
  of $B^+ \to \jpsi K^+ \to \mu^+\mu^- K^+$ decays
  and $B^0_s \to \mu^+\mu^-$ decays that are observed in the MC.
The ratio $\epsilon_{B^+} / \epsilon_{B^0_s}$ is $0.172 \pm 0.015$,
where the sources of uncertainty include the dimuon mass resolution and scale,
the shape of the discriminant distribution, trigger efficiency,
MC statistics, and the shape of the $p_T$ distribution for $B^0_s$ and $B^+$. 
The $B$ meson production ratio was calculated to be
$\frac{f(\overline{b}\to B^+)}{f(\overline{b}\to B^0_s)} = 3.86 \pm 0.54$ 
from the production fractions of Refs.\ \cite{Yao:2006, HFAG:2006}
and the correlation coefficient from Ref.\ \cite{HFAG:2006}.
The branching fractions
    ${\mathcal B}(B^+ \to \jpsi K^+) = (1.008 \pm 0.035)\times 10^{-3}$
and ${\mathcal B}(\jpsi \to \mu^+\mu^-) = 0.0593 \pm 0.0006$
are from Ref.\ \cite{Yao:2006}.
The product of the factors multiplying $n_{B^0_s}$ on the right hand
side of Eq.\ \ref{eq-k} is therefore
$k = {\mathcal B}(B^0_s\to\mu^+\mu^-) / n_{B^0_s} = (1.97 \pm 0.34) \times 10^{-8}$,
often called the single event sensitivity.
The contributions of the various sources of uncertainty to the relative 
uncertainty in $k$ are listed in 
Table \ref{tab-syst}.
  \begin{table}[ht]
  \begin{center}
\begin{tabular}{lr}
\hline
\hline
Source & $\Delta k/k$ \\
\hline
     Mass resolution &  0.007\\
          Mass scale &  0.013\\
Discriminant distribution &  0.030\\
     Trigger efficiency &  0.007\\
       MC statistics &  0.024\\
       $B$ meson $p_T$ spectrum &  0.080\\
$f(\overline{b} \to B^+)/f(\overline{b} \to B^0_s)$ &  0.140\\
${\mathcal B}(B^+ \to \jpsi K^+)$ &  0.035\\
${\mathcal B}(\jpsi \to \mu^+ \mu^-)$ &  0.010\\
 $B^{+}$ fit (stat) &  0.027\\
 $B^{+}$ fit (syst) &  0.022\\
\hline
            Combined &  0.17\\
\hline
\hline
\end{tabular}
  
  \caption{ Sources of uncertainty and their contributions to the
            relative uncertainty in the single event sensitivity
            $k$. \label{tab-syst}}
  \end{center}
  \end{table}

Uncertainties due to differences between the data and MC largely cancel
in the ratio $\epsilon_{B^+} / \epsilon_{B^0_s}$, although not completely.
For instance,
muons from $B^0_s\to\mu^+\mu^-$ decay mostly have higher $p_T$
than muons from $B^+\to \jpsi K^+ \to \mu^+\mu^- K^+$ decay, 
in which the energy is shared among three particles. 
The resulting effect on the efficiency
of the trigger and muon $p_T$ cuts depends on the $p_T$ distribution 
of the parent $B$ mesons, and the shape of this distribution is
the dominant source of uncertainty in $\epsilon_{B^+} / \epsilon_{B^0_s}$.
The extra track in $B^+$ decays together with 
better tracking and vertexing in the MC than in the data
result in an overestimate of $\epsilon_{B^+} / \epsilon_{B^0_s}$ and 
a slight worsening of the limit.
The uncertainty due to modeling of the first four likelihood variables
was estimated to be 3\% based on a comparison between
$B^+$ data and MC.
The uncertainties due to the mass resolution (0.7\%) and scale (1.3\%) 
were estimated by comparing the $\Upsilon(1S) \to \mu^+\mu^-$ 
mass distribution in data and MC. 
Other uncertainties in $\epsilon_{B^+} / \epsilon_{B^0_s}$ are MC 
statistics (2.4\%) and trigger efficiency (0.7\%).


Given two candidates observed in the data, an upper limit 
on $n_{B^0_s}$ was computed
taking into account the expected background 
and uncertainties using a Bayesian method.
The resulting upper limit on the branching fraction is  
${\mathcal B}(B^0_s\to\mu^+\mu^-) <$ \BR95\ at the 95\% C.L.
The expected limit is $0.97\times 10^{-7}$. 
This result improves upon the best previously published upper bound 
for this branching fraction \cite{cdflim}.

%
We thank the staffs at Fermilab and collaborating institutions, 
and acknowledge support from the 
DOE and NSF (USA);
CEA and CNRS/IN2P3 (France);
FASI, Rosatom and RFBR (Russia);
CAPES, CNPq, FAPERJ, FAPESP and FUNDUNESP (Brazil);
DAE and DST (India);
Colciencias (Colombia);
CONACyT (Mexico);
KRF and KOSEF (Korea);
CONICET and UBACyT (Argentina);
FOM (The Netherlands);
Science and Technology Facilities Council (United Kingdom);
MSMT and GACR (Czech Republic);
CRC Program, CFI, NSERC and WestGrid Project (Canada);
BMBF and DFG (Germany);
SFI (Ireland);
The Swedish Research Council (Sweden);
CAS and CNSF (China);
Alexander von Humboldt Foundation;
and the Marie Curie Program.
%

\bibliography{prl.bib}

\begin{thebibliography}{18}
\bibitem[{con()}]{conj}
\bibinfo{note}{Charge conjugate states are included implicitly throughout this
  paper.}
\end{thebibliography}

\end{document}
%